\documentstyle[12pt]{article}
\def\be{\begin{equation}}
\def\ee{\end{equation}}
\def\bea{\begin{eqnarray}}
\def\eea{\end{eqnarray}}
\begin{document}
\begin{center}
{\Large{\bf Moving Mixed Branes in Compact Spacetime }}                  
									      
\vskip .5cm   
{\large Davoud Kamani}
\vskip .1cm
 {\it Institute for Studies in Theoretical Physics and 
Mathematics
\\ Tehran P.O.Box: 19395-5531, Iran}\\
{\it and}\\
{\it  Department of Physics, Sharif University of Technology
P.O.Box 11365-9161}\\
{\sl e-mail: kamani@netware2.ipm.ac.ir}
\\
\end{center}

\begin{abstract} 
In this article we present a general description of two moving branes 
in presence of the $B_{\mu \nu}$ field and gauge fields 
$A^{(1)}_{\alpha_1}$ and $A^{(2)}_{\alpha_2}$ on them, in spacetime in which
some of its directions are compact on tori. Some examples are considered to 
elucidate this general description. Also contribution of the massless
states to the interaction is extracted. Boundary state formalism is a useful
tool for these considerations.
\end{abstract} 
\vskip .5cm

PACS:11.25.-w; 11.25.Mj; 11.30.pb 
\newpage
\section{Introduction}
 
Boundary state formalism, which is a powerful tool for describing the
branes and their interactions, has been successfully applied to a number 
of problems, for example D-branes dynamics 
in different configurations and spacetime dimensions \cite{1,2,3,
4,5,6,7}. On the other hand, back-ground 
fields $B_{\mu \nu}$ and $A_{\alpha}$
(a $U(1)$ gauge field which lives in brane) can be introduced to the
string $\sigma$-model action, to obtain mixed boundary condition (i.e.
a combination of Dirichlet and Neumann boundary conditions) for string
\cite{8}. Previously we obtained the mixed boundary state for a static
mixed brane (i.e. a brane in above back-ground fields), and interaction of
static mixed branes in spacetime in which some of its dimensions
are compactified on tori \cite{8}. We saw that the states emitted from the
branes, which are wrapped around compact directions with internal back-ground 
fields turn out to be dominant along a certain direction. Their windings 
around the compact directions of brane are also correlated with their 
momenta along the brane. 

In addition to the above considerations (i.e. existence of the 
back-ground fields and compactification of spacetime), 
now we consider the motion of the mixed branes.
We will see that the momentum component of the closed string state along
the motion of the brane, is also correlated with its windings around the 
compact directions of the brane. Also back-ground fields, compactification and
velocities all together, cause the interaction amplitude take an interesting 
form. For example when these three exist, the initial position   
$y^{i_0}$ of the brane along the motion appears in the interaction.

In section 2 we obtain the boundary states for moving mixed branes in
compact spacetime.
In section 3 we use of these boundary states to calculate interaction 
of two branes of dimensions $p_1$ and $p_2$ with different internal fields 
${\cal{F}}_1$ and ${\cal{F}}_2$, moving with velocities $V_1$ and $V_2$.
We shall also show that these results reduce to the known cases of the
D-branes in non-compact spacetime. To elucidate our general computations, we
apply our results to special cases: parallel $m_1-m_{1'}$ and perpendicular 
$m_1-m_{1'}$ systems. Finally contribution of the massless
states on the interaction will be obtained. 

In this article a brane in 
back-ground internal fields, is denoted by ``$m_p$-brane'', which is a 
``mixed brane'' with dimension ``$p$''. Since compactification effects
on the interaction of the moving mixed branes do not depend on the fermions
, we will consider only the bosonic string.
  
%%%%%%%%%%%%%%%%%%%%%%%%%%%%%%%%%%%%%%%%%%%%%%%%%%%%%%%%%%%%%%%%%%%%%%%%%%%%%
\section{Moving mixed brane and boundary state} 

  We begin with a $\sigma$-model action containing $B_{\mu\nu}$ field and two  
boundary terms corresponding to the two  $m_{p_{1}}$ and 
$m_{p_{2}}$-branes gauge fields \cite{9}, and their velocities \cite{4,10}

\bea
S &=& -\frac{1}{4\pi \alpha'} \int_{\Sigma} d^2 \sigma
\bigg{(} \sqrt{-g}g^{ab}G_{\mu\nu}
\partial_{a}X^{\mu}\partial_{b}X^{\nu}+\epsilon^{ab}B_{\mu\nu}\partial
_{a}X^{\mu}\partial_{b}X^{\nu}\bigg{)}
\nonumber\\
&~&-\frac{1}{2\pi\alpha'}\int_{({\partial\Sigma})_{1}}
d\sigma \bigg{(} A^{(1)}_{\alpha_1}\partial_{\sigma}X^{\alpha_{1}} + 
V^{i_1}_1 X^0 \partial_{\tau} X^{i_1} \bigg{)}
\nonumber\\
&~&+\frac{1}{2\pi\alpha'}\int_{({\partial\Sigma})_{2}}d\sigma \bigg{(} 
A^{(2)}_{\alpha_2} \partial_{\sigma}X^{\alpha_{2}}+ V^{i_2}_2 X^0                                     
\partial_{\tau} X^{i_2} \bigg{)} \;\;,
\eea
where $\Sigma$ is the world sheet of the closed string exchanged
between the branes. $(\partial\Sigma)
_{1}$ and $(\partial\Sigma)_{2}$ are two boundaries of this world sheet, 
which are at $\tau=0$ and $\tau=\tau_{0}$ respectively .
$A^{(1)}_{\alpha_1}$ and $A^{(2)}_{\alpha_{2}}$ are $U(1)$ gauge 
fields that live in $m_{p_1}$ and $m_{p_2}$-branes. 
$V^{i_1}_1$ and $V^{i_2}_2$ are velocities of the first and the second
branes. The sets $\{{\alpha_1}\}$ and $\{{\alpha_2}\}$ 
 specify the directions on the $m_{p_1}$ and $m_{p_2}$ world volumes, 
 $\{i_1\}$ and $\{i_2\}$ show the directions perpendicular to them. 

 Taking the back-ground fields $G_{\mu\nu}(X)$ and $B_{\mu\nu}(X)$ 
to be constant fields. Vanishing the variation of this action with respect 
 to $X^{\mu}(\sigma,\tau)$ gives the equation of motion of 
 $X^{\mu}(\sigma,\tau)$ and boundary state equations. For the second brane,
 boundary state equations take the form,
\bea
&~&\bigg{(} \partial_{\tau}(X^0- V^{i_2}_2 X^{i_2} )+{\cal{F}}^0
_{(2)\;\beta_2}\partial_
{\sigma}X^{\beta_2} -B^0 \;_{i_2} \partial_{\sigma} ( X^{i_2}
-V^{i_2}_2 X^0 )
\bigg{)}_{\tau=\tau_0}{\mid}B^2_x \;,\tau_0 \rangle = 0 \;,       
\eea
\bea
&~&\bigg{(} \partial_{\tau}X^{\bar{\alpha}_2}+{\cal{F}}^
{\bar{\alpha}_2}_{(2)\;\beta_2}\partial_
{\sigma}X^{\beta_2} -B^{\bar{\alpha}_2}\;_{i_2} \partial_{\sigma} ( X^{i_2}
-V^{i_2}_2 X^0 )
\bigg{)}_{\tau=\tau_0}{\mid}B^2_x \;,\tau_0 \rangle = 0 \;,       
\eea
\bea
&~&{{\delta}(X^{i_{2}} -V^{i_2}_2 X^0)}_{\tau=\tau_0}{\mid}B^2_x \;,\tau_0
\rangle = 0\;\;,
\eea
where $\bar{\alpha}_2$ refers to the spatial directions of the $m_{p_2}$
-brane (i.e. $\bar{\alpha}_2 \neq 0 $), and ${\cal{F}}_2$ 
is total ``field strength'',  
\bea
{\cal{F}}_{(2)\alpha_2\beta_2} \equiv \partial_{\alpha_2} A^{(2)}_{\beta_2} 
-\partial_{\beta_2} A^{(2)}_{\alpha_2}-B_{\alpha_2\beta_2} \;\;.
\eea
The transverse coordinates of the two branes initially are 
$\{y^{i_{1}}_1\}$ and  $\{y^{i_{2}}_2\}$, therefore  
\bea 
&~&\bigg{(} X^{i_2}(\sigma,\tau)-V^{i_2}_2 X^0 (\sigma , \tau)-
y^{i_2}_{2}\bigg{)}_{\tau = \tau_0}{\mid}B^2_x \;,\tau_0 \rangle = 0 \;\;.
\eea   
This implies $\partial_{\sigma} ( X^{i_2} - V^{i_2}_2 X^0 )$ 
vanish on the boundary and be dropped from the equations (2) and (3).

Solution of the equation of motion of the closed string is
\bea
X^{\mu}(\sigma,\tau)= x^{\mu}+2{\alpha'}p^{\mu}\tau+2L^{\mu}\sigma+
 \frac{i}{2}\sqrt{2\alpha'}\sum_{m\neq 0}\frac{1}{m}(\alpha^{\mu}_{m}
 e^{-2im(\tau-
 \sigma)}+\tilde\alpha^{\mu}_{m}e^{-2im(\tau+\sigma)}) \;\; ,
\eea
where $L^{\mu}$ is zero for non-compact directions, for compact 
directions we have $L^{\mu} = N^{\mu}R^{\mu}$ and $p^{\mu} = \frac{M^{\mu}}
{R^{\mu}}$ , in which $N^{\mu}$ is the winding number and $M^{\mu}$ is the 
momentum number of the closed string state, also $R_{\mu}$ is the radius 
of compactification in the compact direction $X^{\mu}$. 

 Combining the solution of the equation of motion and the 
boundary state equations, assuming non-compact time direction,
we obtain the boundary state equations 
in terms of modes,
\bea
\bigg{[} \bigg{(} \alpha^0 _m - V^{i_2}_2 \alpha^{i_2}_m 
-{\cal{F}}^0_{(2) \;\;\;{\bar{\beta}_2}} \alpha^{\bar{\beta}_2}_m \bigg{)}
e^{-2im\tau_{0}}
+ \bigg{(} \tilde{\alpha} ^0 _{-m} - V^{i_2}_2 
\tilde{\alpha}^{i_2} _{-m} + {\cal{F}}^0 _{(2)\;\;\;\bar{\beta}_2}
\tilde\alpha^{\bar{\beta}_2}_{-m}\bigg{)}e^{2im\tau_0}
 \bigg{]} \mid B^2_x,\tau_0 \rangle = 0 \;\;,
\eea
\bea
\bigg{[} \bigg{(} \alpha^{\bar{\alpha}_2} _m  
-{\cal{F}}^{\bar{\alpha}_2}_{(2) \;\;\;\beta_{2}} \alpha^{\beta_2}_m \bigg{)}
e^{-2im\tau_{0}}+ \bigg{(} \tilde{\alpha} ^{\bar{\alpha}_2} _{-m} 
+ {\cal{F}}^{\bar{\alpha}_2} _{(2)\;\;\;\beta_2}
\tilde\alpha^{\beta_2}_{-m}\bigg{)}e^{2im\tau_0}
 \bigg{]} \mid B^2_x,\tau_0 \rangle = 0 \;\;,
\eea
\bea
\bigg{[}\bigg{(}\alpha^{i_2}_{m} - V^{i_2}_2 \alpha^0_m \bigg{)}
e^{-2im\tau_{0}}-\bigg{(} \tilde\alpha^{i_2}_{-m} - V^{i_2}_2 \tilde{\alpha}
^0_{-m}\bigg{)}e^{2im{\tau_0}}\bigg{]}\mid B^2_x,\tau_{0} \rangle = 0 \;\;,
\eea
for the oscillating part, and

\bea
\bigg{(}p^0 -V^{i_2}_2 p^{i_2} + \frac{1}{\alpha'}{\cal{F}}^0
_{(2)\;\bar{\beta}_2} L^{\bar{\beta}_{2}}
\bigg{)}_{op} \mid B^2_x,\tau_{0} \rangle = 0 \;\;,
\eea                                           
\bea
\bigg{(}p^{\bar{\alpha}_2}+\frac{1}{\alpha'}{\cal{F}}^{\bar{\alpha}_2}
_{(2)\;\bar{\beta}_2} L^{\bar{\beta}_2}
\bigg{)}_{op} \mid B^2_x,\tau_{0} \rangle = 0 \;\;,
\eea                                           
\bea
\bigg{(}x^{i_2}-V^{i_2}_2 x^0 - y^{i_2}_{2} +2{\alpha'} \tau_{0} ( p^{i_2}
-V^{i_2}_2 p^0 ) \bigg{)}_{op}
{\mid}B^2_x,\tau_{0} \rangle = 0 \;\; ,
\eea
\bea                
L^{i_2}\mid B^2_x,\tau_{0} \rangle=0 \;\; ,
\eea
for the zero mode part. The oscillating part can be written as,  
\bea
\bigg{(} \alpha^{\mu}_m e^{-2im\tau_0} + S^{\mu} _{\;\;\nu} \tilde{\alpha} 
^{\nu}_{-m} e^{2im\tau_0} \bigg{)} \mid B^2_x \;,\tau_0 \rangle = 0\;\; ,
\eea
\bea
S \equiv M^{-1} N\;\;,
\eea
where matrices $M$ and $N$, which depend on ${\cal{F}}_2$ and $V_2$ are
defined by
\bea
\left \{ \begin{array}{cl} 
M^0_{\;\;\;\mu} = \delta^0_{\;\;\;\mu} - V^{i_2}_2 \delta^{i_2}_{\;\;\;\mu}
-{\cal{F}}^0 _{(2)\beta_2} \delta^{\beta_2}_{\;\;\;\mu}\\
M^{\bar{\alpha}_2}_{\;\;\;\;\mu} = \delta^{\bar{\alpha}_2}_{\;\;\;\;\mu} 
-{\cal{F}}^{\bar{\alpha}_2} _{(2)\;\;\;\beta_2} 
\delta^{\beta_2}_{\;\;\;\mu} \\
M^{i_2}_{\;\;\;\mu} = \delta^{i_2}_{\;\;\;\mu} - V^{i_2}_2 
 \delta^0_{\;\;\;\mu}
 \end{array} \right.
\eea
and
\bea
\left \{ \begin{array}{cl}
N^0_{\;\;\;\mu} = \delta^0_{\;\;\;\mu} - V^{i_2}_2 \delta^{i_2}_{\;\;\;\mu}
+{\cal{F}}^0 _{(2)\;\;\;\beta_2} \delta^{\beta_2}_{\;\;\;\mu}\\
N^{\bar{\alpha}_2}_{\;\;\;\;\mu} = \delta^{\bar{\alpha}_2}_{\;\;\;\;\mu} 
+{\cal{F}}^{\bar{\alpha}_2} _{(2)\;\;\;\beta_2} 
\delta^{\beta_2}_{\;\;\;\mu} \\
N^{i_2}_{\;\;\;\mu} = -\delta^{i_2}_{\;\;\;\mu} + V^{i_2}_2 
\delta^0_{\;\;\;\mu}
\end{array} \right.
\eea
These definitions of the matrices $M$ and $N$ imply $S$ be an orthogonal
matrix, i.e. $MM^T = NN^T$, one can investigate this identity element by
element.

We now extract the boundary state from the equations (11-15). Oscillators in
(15) results in
\bea
\mid B^2_{osc} \rangle =  exp \bigg{[}-\sum_{m=1}^{\infty}\bigg{(}\frac{1}{m}
e^{4im\tau_0}\alpha^{\mu}_{-m}S^{(2)}_
 {\mu\nu}{\tilde\alpha}^{\nu}_{-m}\bigg{)} \bigg{]}\mid 0 \rangle \;\;.
\eea
From now on, we restrict ourselves to the case that $m_{p_1}$ and $m_{p_2}$
-branes move along the $x^{i_0}$-direction which is perpendicular to the both
of them, therefore $V^{i_0}_1 \equiv V_1$ and $V^{i_0}_2 \equiv V_2$ and
all other components of the velocities are zero. These imply the 
solutions of the zero mode part to be as the following,
\bea
 \mid B^2_x , \tau_{0} \rangle ^{(0)} = \sum_{\{{p^{\alpha_2}}\}}
 \mid B^2_x , \tau_0\;,\;p^{\alpha_2}  \rangle ^{(0)} \;\;,
\eea
\bea  
\mid B^2_x , \tau_0 , p^{\alpha_2}  \rangle ^{(0)}
&=& \frac{T_{p_2}}{2}
\sqrt{det M_2} \;exp \bigg{[} i\alpha'\tau_0 \bigg{(}
\gamma^2_2(p^{i_0}_{op}- V_2p^0_{op})^2 
+\sum_{j_2 \neq i_0}(p^{j_2}_{op})^2 \bigg{)} \bigg{]}
\nonumber\\
&~&\times \delta(x^{i_0}-V_2x^0-y^{i_0}_2) 
 \prod_{j_2 \neq i_0} \delta(x^{j_2}-y^{j_2}_2) 
\prod_{\bar{\alpha}_2}\mid p^{\bar{\alpha}_2} \rangle 
\nonumber\\ 
&~&\times \prod_{j_2 \neq i_0}\mid p^{j_2}_L = p^{j_2}_R = 0 \rangle
\mid p^0 \rangle \mid p^{i_0}_L = p^{i_0}_R = \frac{1}{2}V_2p^0
\rangle
\eea  
where $\gamma_2$ is $1/ \sqrt{1-V^2_2} $ and $T_{p_2}$ is the $D_{p_2}$
-brane tension \cite{11}. Path integral with boundary action gives 
$\sqrt{det M_2}$ \cite{12,13,14}, and for the ${\cal{F}}_2 = 0$ it becomes 
$\frac{1}{\gamma_2}$. In this state momentum components are,
\bea
p^{\bar{\alpha}_2} = -\frac{1}{\alpha'} {\cal{F}}^{\bar{\alpha}_2}
_{(2)\;\;\;\beta_{2c}}{\ell}^{\beta_{2c}}\;\;,
\eea
\bea
p^0 = -\frac{\gamma^2_2}{\alpha'} {\cal{F}}^0_{(2)\;\;\;\beta_{2c}}{\ell}
^{\beta_{2c}} \;\;,
\eea
\bea
p^{i_0} = -\frac{V_2\gamma^2_2}{\alpha'} {\cal{F}}^0
_{(2)\;\;\;\beta_{2c}}{\ell}^{\beta_{2c}}\;\;,
\eea
therefore, for the closed string state emitted from the 
moving brane with back-ground fields in compact spacetime, 
besides, that the momentum components along the world volume of the brane are 
non-zero and are quantized, the momentum component 
along the motion of the brane is also non-zero and is 
quantized. More details of (22-24) for $V_2 = 0$, 
can be found in \cite{8}. In
(20), due to the relations (22-24),
the summation over the momentum components can be changed to a sum over
winding numbers, $\{N^{\alpha_{2c}}\}$.

Ghost part of the boundary state has the form
\bea
  \mid B_{gh}, \tau_{0} \rangle = exp \bigg{[} 
  \sum_{m=1}^{\infty}{e^{4im\tau_0}}
  (c_{-m}{\tilde{b}}_{-m}-b_{-m}{\tilde{c}}_{-m})\bigg{]}\frac{c_0+
  \tilde{c}_0}{2}\mid q=1\rangle
  \mid \tilde{q}=1 \rangle
\eea
 
 %%%%%%%%%%%%%%%%%%%%%%%%%%%%%%%%%%%%%%%%%%%%%%%%%%%%%%%%%%%%%%%%%%%%%%%%%%%
\section{Moving mixed branes interaction}

Before calculation of the interaction amplitude, let us introduce some  
notations for the positions of these two mixed branes. The set $\{\bar{i}\}$
shows the directions perpendicular to the both of the branes, in which
$i_0$ is not in $ \{ \bar{i}\}$, the set $\{\bar{u}\}$ 
for the directions along the 
both of them, in which $0$ is not in $\{ \bar{u} \}$, the 
set $\{\alpha'_1\}$ for the 
directions along the $m_{p_1}$ and perpendicular to the $m_{p_2}$, and the 
set $\{ \alpha'_2\}$ for the directions along the $m_{p_2}$ and perpendicular
to the $m_{p_1}$-branes. It can be seen that for example 
\bea
\{i_1\} = \{\bar{i}\} \bigcup \{i_0\} \bigcup \{\alpha'_2\} \;\;,\;\;
\{\alpha_1\} = \{\alpha'_1\} \bigcup \{\bar{u}\} \bigcup \{0\} \;\;\;.
\eea

The complete boundary state can be written as the product $
\mid B \rangle = \mid B_{x} \rangle \mid B_{gh} \rangle $,
therefore the interaction amplitude is
\bea 
{\cal{A}} = \langle B_1\mid D \mid B_2,\tau_0 =0 \rangle \;\;,
\eea
where ``$D$'' is the closed string propagator. The final result is

\bea
{\cal{A}} &=& \frac{T_{p_1}T_{p_2}}{4(2\pi)^{\bar{d_i}}}
\frac{\alpha'\gamma_1 \gamma_2}{sinh \omega} 
\sqrt{det M_1 detM_2} 
\int_0^{\infty} dt \bigg{\{} e^{4at}  
\nonumber\\
&~&\times \prod_{n=1}^\infty \bigg{(}[ det(1-S_1S_2^T 
e^{-4nt})]^{-1}(1-e^{-4nt})^2 \bigg{)}
\nonumber\\
&~&\times \bigg( \sqrt{\frac{\pi}{\alpha't}} \; \bigg)^{\bar{d}_{i_n}}
e^{ -\frac{1}{4\alpha't}\sum_{\bar{i}_n}(y^{\bar{i}_n}_1-y^{\bar{i}_n}_2)^2 }
\prod_{\bar{i}_c}\Theta_3 \bigg( \frac{y^{\bar{i}_c}_1 
- y^{\bar{i}_c}_2 }{2\pi R_{\bar{i}_c}} \mid 
\frac{i\alpha't}{\pi (R_{\bar{i}_c})^2}\bigg)
\nonumber\\
&~& \times \sum_{\{N^{u_c}\}} \bigg{[} (2\pi)^{\bar{d}_u}
[\prod_{\bar{u}}\delta(p^{\bar{u}}_1 - p^{\bar{u}}_2)]
\; exp [\frac{i}{\alpha'}{\ell}^{u_c}
\bigg{(}{\cal{F}}^{\alpha'_1}_{(1)\;\;u_c} y^{\alpha'_1}_{2}-
{\cal{F}}^{\alpha'_2}_{(2)\;\;u_c}y^{\alpha'_2}_{1}
\nonumber\\
&~& +\phi_{u_c}(12)y^{i_0}_2 - \phi_{u_c}(21)y^{i_0}_1 \bigg{)} ]
 exp [-\frac{t}{\alpha'}{\ell}^{u_c}{\ell}^{v_c}
\bigg{(}\delta_{u_cv_c}+f^{(+)}_{u_c}f^{(-)}_{v_c}
\nonumber\\
&~& +{\cal{F}}^{\bar{u}}
_{(1)\;\;u_c}{\cal{F}}_{(2)\;\;\bar{u}v_c}+
{\cal{F}}^{\alpha'_1}_{(1)\;\;u_c}{\cal{F}}^{\alpha'_1}_{(1)\;\;v_c}+
{\cal{F}}^{\alpha'_2}_{(2)\;\;u_c}{\cal{F}}^{\alpha'_2}_{(2)\;\;v_c}\bigg{)}] 
\bigg{]} \;\bigg{\}}
\eea
where $a,\; \omega,\; \phi_{u_c} (12),\; $ and $f^{(+)}_{u_c}$ are
\bea
a= (d-2)/24 \;\;\;,\; \omega = \mid \omega_2-\omega_1 \mid\;\;\;\;,
\;\;V_{1,2} = tgh \omega_{1,2}\;\;,
\eea
\bea
\phi_{u_c}(12) = \frac{1}{V_2-V_1} \bigg{[} \gamma^2_2(1+V_1V_2) {\cal{F}}^0
_{(2)\;\;\;u_c} - \gamma^2_1 (1+V^2_1) {\cal{F}}^0_{(1)\;\;\;u_c} 
\bigg{]}\;\;\;,
\eea
\bea
f^{(+)}_{u_c} = \frac{1}{\mid V_1-V_2\mid } \bigg{[} \gamma^2_2(1+V_1) 
(1+V^2_2){\cal{F}}^0 _{(2)\;\;\;u_c} - \gamma^2_1(1+V_2)(1+V^2_1) 
{\cal{F}}^0_{(1)\;\;\;u_c} \bigg{]}\;\;\;,
\eea
and $\phi_{u_c}(21)$ is given 
in (31) with the exchange $1 \leftrightarrow 2$
, also for $f^{(-)}_{u_c}$, change the signs of $V_1$ and $V_2$ in (32). 

In this amplitude $p^{\bar{u}}_1 = -\frac{1}{\alpha'} {\cal{F}}^{\bar{u}}
_{(1)\;\;\;v_c}N^{v_c}R^{v_c}$ and $p^{\bar{u}}_2 = -\frac{1}{\alpha'} 
{\cal{F}}^{\bar{u}} _{(2)\;\;\;v_c}N^{v_c}R^{v_c}$. Indices $\{u_c , v_c\}$ 
show the compact part of $\{\bar{u}\}$, also $\bar{d}_i$ and 
$\bar{d}_{i_n}$ show
the dimensions of $\{X^{\bar{i}}\}$ and $\{X^{\bar{i}_n}\}$ respectively.
The sets $\{\bar{i}_n\}$ and $\{\bar{i}_c\}$ show the 
non-compact and compact part
of $\{\bar{i}\}$. Under the exchange of indices ``1'' and ``2'' this 
amplitude is symmetric i.e. ${\cal{A}}_{(1,2)}={\cal{A}}^* _{(2,1)}$
, as expected (see(28)). From (29) and (31) 
we see that the non-zero
electric fields, $E^{(1)}_{u_c} = {\cal{F}}_{(1)\;\;0 u_c}$ and $E^{(2)}
_{u_c} = {\cal{F}}_{(2)\;\;0 u_c}$, spacetime 
compactification and motion of the branes 
cause the $y^{i_0}_1$ and $y^{i_0}_2$ to appear in the interaction.

The momentum delta functions put some restrictions on the summation. The 
term corresponding to $N^{u_c} =0 $ for all $u_c$, gives $p^{\bar{u}}_1 = 
p^{\bar{u}}_2 = 0$, and is always present. Other terms occur only if the
two internal back-ground fields and radii of compactification with some
sets $\{N^{u_c}_s\}$ satisfy the relation $\sum_{v_c}(
{\cal{F}}^{\bar{u}}_{(1)\; v_c}N^{v_c}_s R^{v_c}) = 
\sum_{v_c}({\cal{F}}^{\bar{u}}_{(2)\;v_c} N^{v_c}_s R^{v_c})$ for all 
$\bar{u}$. In this case common volume of the branes 
$(V_{\bar{u}})$ explicitly appears in the amplitude, therefore
\bea
{\cal{A}} &=& \frac{T_{p_1}T_{p_2}V_{\bar{u}}}{4(2\pi)^{\bar{d_i}}}
\frac{\alpha'\gamma_1 \gamma_2}{sinh \omega} 
\sqrt{det M_1 detM_2} 
\int_0^{\infty} dt \bigg{\{} e^{4at} 
\nonumber\\
&~&\times  \prod_{n=1}^\infty \bigg{(}[ det(1-S_1S_2^T 
e^{-4nt})]^{-1}(1-e^{-4nt})^2 \bigg{)}
\nonumber\\
&~&\times \bigg( \sqrt{\frac{\pi}{\alpha't}} \; \bigg)^{\bar{d}_{i_n}}
e^{ -\frac{1}{4\alpha't}\sum_{\bar{i}_n}(y^{\bar{i}_n}_1-y^{\bar{i}_n}_2)^2 }
\prod_{\bar{i}_c}\Theta_3 \bigg( \frac{y^{\bar{i}_c}_1 
- y^{\bar{i}_c}_2 }{2\pi R_{\bar{i}_c}} \mid 
\frac{i\alpha't}{\pi (R_{\bar{i}_c})^2}\bigg)
\nonumber\\
&~& \times \bigg{[} \bigg{[} 1 + \sum_s \{ 
 exp \bigg{[}-\frac{t}{\alpha'}{\ell}^{u_c}_s{\ell}^{v_c}_s
\bigg{(}\delta_{u_cv_c}+f^{(+)}_{u_c}f^{(-)}_{v_c}
+{\cal{F}}^{\bar{u}}
_{(1)\;\;u_c}{\cal{F}}_{(2)\;\;\bar{u}v_c}+
{\cal{F}}^{\alpha'_1}_{(1)\;\;u_c}{\cal{F}}^{\alpha'_1}_{(1)\;\;v_c}
\nonumber\\
&~&+{\cal{F}}^{\alpha'_2}_{(2)\;\;u_c}{\cal{F}}^{\alpha'_2}_{(2)\;\;v_c}
\bigg{)} \bigg{]} exp \bigg{[}\frac{i}{\alpha'}{\ell}^{u_c}_s
\bigg{(}{\cal{F}}^{\alpha'_1}_{(1)\;\;u_c} y^{\alpha'_1}_{2}-
{\cal{F}}^{\alpha'_2}_{(2)\;\;u_c}y^{\alpha'_2}_{1}
+\phi_{u_c}(12)y^{i_0}_2 - \phi_{u_c}(21)y^{i_0}_1 
\bigg{)} \bigg{]} \} \bigg{]}\bigg{]} 
\bigg{\}} \;\;\; ,
\eea
where ${\ell}^{u_c}_s = N^{u_c}_s R^{u_c} $. If there are no sets 
$\{N^{u_c}_s\}$ then $[[ \;\;\;\; ]] = 1$.
For parallel mixed branes with the same dimension those terms containing
$\alpha'_1$ and $\alpha'_2$ disappear.

The all effects of compactification are in the last bracket and the 
products of $\Theta_3$-functions, therefore amplitude in non-compact 
spacetime is
\bea
{\cal{A}}_{(nc)} &=& \frac{T_{p_1}T_{p_2}V_{\bar{u}}}{4(2\pi)^{\bar{d}_i}}
\frac{\alpha'\gamma_1 \gamma_2}{sinh \omega} 
\sqrt{det M_1 detM_2}
\int_0^{\infty} dt \bigg{\{}  
\bigg( \sqrt{\frac{\pi}{\alpha't}} \; \bigg)^{\bar{d}_i}
e^{ -\frac{1}{4\alpha't}\sum_{\bar{i}}(y^{\bar{i}}_1-y^{\bar{i}}_2)^2 }
\nonumber\\
&~& \times e^{4at} \prod_{n=1}^\infty \bigg{(}[ det(1-S_1S_2^T 
e^{-4nt})]^{-1}(1-e^{-4nt})^2 \bigg{)}\;\;.
\eea
For parallel $D_p$-branes (i.e. ${\cal{F}}_1 = {\cal{F}}_2 = 0$ ) along 
$(X^1,X^2,...,X^p)$ with $T_p=\frac{\sqrt{\pi}}{2^{(d-10)/4}} (4 \pi^2 
\alpha' )^{(d-2p-4)/4}$ and $t \rightarrow \pi t/2$, the amplitude
${\cal{A}}_{(nc)}$ reduces to result of \cite{5}.
%%%%%%%%%%%%%%%%%%%%%%%%%%%%%%%%%%%%%%%%%%%%%%%%%%%%%%%%%%%%%%%%%%%%%%%%%%%%%
\section{Examples}
To elucidate these general computations, we apply our results to 
special cases.
These are: parallel $m_1-m_{1'}$ branes along $X^1$-direction, moving
along $X^2$-direction and perpendicular $m_1-m_{1'}$ branes along $X^1$
and $X^2$ directions, moving along $X^3$-direction. For both of these examples
we give the following amplitude,
\bea
{\cal{A}}_{(1,1')} &=& \frac{T^2_1 L}{4(2\pi)^{d-r}}
\frac{\alpha'\gamma_1 \gamma_2}{sinh \omega} 
\sqrt{(1-E^2_1-V^2_1)(1-E^2_2-V^2_2)}
\int_0^{\infty} dt \bigg{\{}  
\nonumber\\
&~&\times e^{4at} \prod_{n=1}^\infty \bigg{(}[ det(1-H_1H_2^T 
e^{-4nt})]^{-1}(1-e^{-4nt})^{2+r-d} \bigg{)}
\nonumber\\
&~&\times \bigg( \sqrt{\frac{\pi}{\alpha't}} \; \bigg)^{\bar{d}_{i_n}}
e^{ -\frac{1}{4\alpha't}\sum_{\bar{i}_n}(y^{\bar{i}_n}_1-y^{\bar{i}_n}_2)^2 }
\prod_{\bar{i}_c}\Theta_3 \bigg( \frac{y^{\bar{i}_c}_1 
- y^{\bar{i}_c}_2 }{2\pi R_{\bar{i}_c}} \mid 
\frac{i\alpha't}{\pi (R_{\bar{i}_c})^2}\bigg)
\nonumber\\
&~& \times \theta(t,R,V,{\cal{F}}) \bigg{\}} \;\;.
\eea
{\bf parallel ${\bf m_1}$-branes }

For this system we have $L=2\pi R_1$, $r=3 , E_1={\cal{F}}_{(1)\;01} , 
E_2={\cal{F}}_{(2)\;01} , {\bar{i}} \in \{ 3,4,...,d-1\}$ and function 
$\theta$ is
\bea
\theta(E_1,E_2,V_1,V_2,t,R_1)=\Theta_3 \bigg{(} \frac{(\phi_{12}y^2_2
-\phi_{21}y^2_1)R_1}{2\pi \alpha'} \mid \frac{it(1+f_{+}f_{-})R^2_1}{\pi
\alpha'} \bigg{)} \;\;,
\eea
where $\phi_{12}$ and $f_{+}$ are the same as in (31) and (32) 
with $u_c=1$, and
corresponding expressions for $\phi_{21}$ and $f_{-}$ . If $X^1$-direction 
is not compact then $\theta=1$, and therefore interaction will be independent
to $y^2_1 , y^2_2 , \phi_{12} $ and $f_{\pm}$. Also matrices $H_1$ and
$H_2$ have similar form, for simplicity drop the indices 1 and 2, therefore
\bea
H^p_{\;\;q} =\frac{1}{1-E^2-V^2} \left( \begin{array}{ccc}
1+E^2+V^2 & -2E & -2V \\
-2E & 1+E^2-V^2 & 2EV \\
2V & -2EV & -(1-E^2+V^2)
\end{array} \right) \;\;,
\eea
where $p,q=0,1,2$. Note that for this system matrices 
$S_1$ and $S_2$ have the common
form
\bea
S^{\mu}_{\;\;\nu} = \left( \begin{array}{cc}
H^p_{\;\;q} & 0\\
0 & -{\bf 1}_{(d-3) \times (d-3)}
\end{array} \right)\;\;.
\eea
In this example the matrix $S_{\mu \nu} = 
\eta_{\mu \lambda}S^{\lambda}_{\;\;\;\nu}$
is exactly that, which is given in Ref.\cite{4}, with notation 
$(\eta.\Lambda.T)_{\mu \nu}$, (our definition of $E$ is negative of 
\cite{4}).

{\bf perpendicular ${\bf m_1}$-branes}

For this system there are $L=1,r=4,E_1={\cal{F}}_{(1)\;01},E_2=
{\cal{F}}_{(2)\;02},\bar{i} \in \{4,5,...,d-1\},\theta(t,R,V,{\cal{F}})=1 $
and matrices $H_1$ and $H_2$ are
\bea
H_2 = \frac{1}{1-E^2_2-V^2_2} \left( \begin{array}{cccc}
1+E^2_2+V^2_2 & 0 & -2E_2 & -2V_2\\
0 & -(1-E^2_2-V^2_2) & 0 & 0 \\
-2E_2 & 0 & 1+E^2_2-V^2_2 & 2E_2V_2\\
2V_2 & 0 &-2E_2V_2 & -(1-E^2_2+V^2_2)
\end{array} \right) \;\;,
\eea
and matrix $H_1$ can be obtained from the $H_2$ as the following: 
change the second and
third columns with each other and again in this new matrix change the second
and third rows with each other, finally change the index ``2'' to ``1''.
%%%%%%%%%%%%%%%%%%%%%%%%%%%%%%%%%%%%%%%%%%%%%%%%%%%%%%%%%%%%%%%%%%%%%%%%%%%%
\section{Massless states contribution to the amplitude}
In this part, to see how distant branes interact 
we obtain the interaction of these 
branes due to the exchange of the massless states.
As the metric, antisymmetric tensor
and dilaton states have zero winding and zero momentum numbers, only
the term with $N^{u_c}=0$ (for all $u_c$) corresponds to these three massless
states. By using the identity $detA=e^{Tr(lnA)}$ for a matrix $A$, there is 
the following limit for $d=26$,
\bea
\lim _{q \rightarrow 0}\; \frac{1}{q}\; \prod_{n=1}
^{\infty} \bigg{[} (1-q^n)^2 [det(1-S_1S^T_2q^n)]^{-1} \bigg{]}
= \lim_{q \rightarrow 0}\; \frac{1}{q}\; +\;\bigg{(} Tr(S_1S^T_2)\; - \; 2 
\bigg{)} \;\; ,
\eea
where $q=e^{-4t}$, putting out the tachyon divergence, contribution of these 
three massless states becomes
\bea
{\cal{A}}_0 &=& \frac{T_{p_1}T_{p_2}V_{\bar{u}}}{4(2\pi)^{\bar{d}_i}}
\frac{\alpha'\gamma_1 \gamma_2}{sinh \omega} 
\sqrt{det M_1 detM_2}\;[Tr(S_1S^T_2)-2]
\int_0^{\infty} dt \bigg{[} \bigg( \sqrt{\frac{\pi}{\alpha't}} \; 
\bigg)^{\bar{d}_{i_n}} 
\nonumber\\
&~&\times e^{ -\frac{1}{4\alpha't}\sum_{\bar{i}_n}(y^{\bar{i}_n}_1
-y^{\bar{i}_n}_2)^2} \prod_{\bar{i}_c}\Theta_3 \bigg( \frac{y^{\bar{i}_c}_1 
- y^{\bar{i}_c}_2 }{2\pi R_{\bar{i}_c}} \mid 
\frac{i\alpha't}{\pi (R_{\bar{i_c}})^2}\bigg) \bigg{]}\;\;.
\eea
We see that integrand completely comes from the directions perpendicular
to the both of the branes (except the direction of motion 
$X^{i_0}$), it also is independent
of the fields and velocities of the branes. For the parallel $m_p$-branes 
in non-compact spacetime, this reads as
\bea
{\cal{A}}_0 = \frac{T^2_p}{4} \frac{\gamma_1 \gamma_2 V_p}
{sinh \omega} \sqrt{det M_1 detM_2}\;[Tr(S_1S^T_2)-2]
G_{24-p}(\bar{Y} ^2) \;\;,
\eea
where $\bar{Y}^2 = \sum_{\bar{i}} (y^{\bar{i}}_1 - y^{\bar{i}}_2)^2$ is 
impact parameter.

%%%%%%%%%%%%%%%%%%%%%%%%%%%%%%%%%%%%%%%%%%%%%%%%%%%%%%%%%%%%%%%%%%%%%%%%%%%%
\section{Conclusion}
					 
We explicitly showed that how total field strength, velocity of the brane and 
compactification effects appear in the boundary state. These cause 
the closed string state emitted from the brane 
to have a quantized momentum along the brane and,
along the motion of the brane.

Interaction amplitude takes the general form under the influence of total 
field strengths $({\cal{F}}_1 , {\cal{F}}_2)$, velocities $(V_1 , V_2)$,
dimensions $(p_1 , p_2)$ and compactification. In non-compact 
spacetime exchange of the massless states between the parallel $m_p$-branes
depends on the impact parameter as $1/\mid {\bar{Y}} \mid ^{(22-p)}$.

The formalism can be extended to include type IIA and type IIB superstring 
theories. 

{\bf Acknowledgement:} 
The author would like to thank H. Arfaei for fruitful discussion and also 
from M.M. Sheikh-Jabbari and A.H. Fatollahi
for reading the manuscript.

%%%%%%%%%%%%%%%%%%%%%%%%%%%%%%%%%%%%%%%%%%%%%%%%%%%%%%%%%%%%%%%%%%%%%%%%%%%%

\end{document}